\documentstyle[pictex,12pt]{article}

\newcommand{\beq}{\begin{equation}}
\newcommand{\eeq}{\end{equation}}
\newcommand{\beqa}{\begin{eqnarray}}
\newcommand{\eeqa}{\end{eqnarray}}
\newcommand{\nn}{\nonumber}
\newcommand{\eq}[1]{(\ref{#1})}

\newcommand{\alimit}{\stackrel{\alpha' \rightarrow 0}{\longrightarrow}}
\newcommand{\ra}{\rightarrow}
\newcommand{\zalmu}{z_{\alpha_{\mu}}}
\newcommand{\zbemu}{z_{\beta_{\mu}}}
\newcommand{\zai}{z_{\alpha_i}}
\newcommand{\zbi}{z_{\beta_i}}
\newcommand{\zaj}{z_{\alpha_j}}
\newcommand{\zbj}{z_{\beta_j}}
\newcommand{\imtau}[1]{(2\pi\mbox{Im}\tau)^{-1}_{#1}}
\newcommand{\NP}[1]{ {\it Nucl.~Phys.} {\bf #1}}
\newcommand{\PL}[1]{ {\it Phys.~Lett.} {\bf #1}}

\newcommand{\PR}[1]{ {\it Phys.~Rev.} {\bf #1}}
\newcommand{\PRL}[1]{ {\it Phys.~Rev.~Lett.} {\bf #1}}

\begin{document}
\topmargin 0pt
\oddsidemargin 1mm
\begin{titlepage}
\begin{flushright}
NBI-HE-96-19\\
hep-th/9604152\\
April, 1996\\
\end{flushright}
\setcounter{page}{0}
\vspace{15mm}
\begin{center}
{\Large   MULTILOOP WORLD-LINE GREEN FUNCTIONS \\
FROM STRING THEORY } 
\vspace{20mm}

{\large Kaj 
Roland~\footnote{Supported by the Carlsberg Foundation,\,\,\,
roland@nbivms.nbi.dk}} 
{\sc and } 
{\large Haru-Tada Sato~\footnote{Fellow 
of the Danish Research Academy,\,\,\,  sato@alf.nbi.dk}}\\

{\em The Niels Bohr Institute, University of Copenhagen\\
     Blegdamsvej 17, DK-2100 Copenhagen, Denmark}\\
\end{center}
\vspace{7mm}

\begin{abstract}
We show how the multiloop bosonic Green function 
of closed string theory reduces to the world-line Green function 
as defined by Schmidt and Schubert in the limit where the string 
world-sheet degenerates into a $\Phi^3$ particle diagram.
To obtain this correspondence we have to make an appropriate 
choice of the local coordinates defined on the degenerate string world 
sheet. We also present a set of simple rules that specify, 
in the explicit setting of the Schottky parametrization, which is the 
corner of moduli space corresponding to a given multiloop $\Phi^3$ 
diagram.

\end{abstract}

\vspace{1cm}

\end{titlepage}
\newpage
\renewcommand{\thefootnote}{\arabic{footnote}}

\section{Introduction}
\setcounter{equation}{0}
\indent

The relation between string theory and field theory has been 
investigated in considerable detail over the last few years. 
Any string theory reduces to an effective 
field theory in the limit where the inverse string tension 
$\alpha' \ra 0$. But string theory also manages to organize the 
scattering amplitudes in a very compact form, and this makes the
detailed investigation of the field theory limit both non-trivial 
and potentially very useful.

For example, Bern and Kosower obtained a set of simple rules for the 
calculation of one-loop amplitudes in pure Yang-Mills theory 
by systematically analyzing the field theory limit of a suitable 
heterotic string theory \cite{BK}. These rules involved only 
$\Phi^3$-like particle diagrams and provided a very significant 
improvement in computational efficiency over traditional Feynman diagram 
techniques. In terms of field theory the rules were subsequently 
seen to correspond to an ingenious combination of background field 
gauge and Gervais-Neveu gauge~\cite{berndunbar}. 
Bern-Kosower rules were also obtained for gravity \cite{gra}, where 
the improvement over traditional techniques was
even more spectacular.

Later on, Strassler obtained the Bern-Kosower rules 
directly from a first-quantized world-line formulation of one-loop 
Yang-Mills theory \cite{Str}. This particle approach has been applied
to several other theories at the one-loop level~\cite{others} 
and has also been generalized to 
multiloop $\Phi^3$-theory \cite{SSphi} and QED 
\cite{SSqed} by Schmidt and 
Schubert and to multiloop scalar QED by Daikouji, Shino 
and Sumino \cite{sumino}.

It would be interesting to find multiloop generalizations of the 
Bern-Kosower rules also for Yang-Mills theory.
To solve this problem it seems very likely that one will need some 
input from string theory --after all, the subtle combination of 
gauge choices inherent in the one-loop Bern-Kosower rules 
would have been quite hard to discover without the help of string 
theory. On the other hand, a direct generalization of the 
Bern-Kosower analysis to more than one loop is not 
straightforward~\cite{me},\cite{me2}, due to the complicated nature 
of the (super) moduli space of higher genus Riemann surfaces. 
From this point of view the multiloop world-line formalism seems 
the more promising one.
Thus, both the string approach of Bern and Kosower and the particle 
approach of Strassler may be useful, each in its own way.

In this paper we bring the two points of view closer together by 
showing how the multiloop bosonic world-line Green functions of 
particle theory, as defined by Schmidt and Schubert \cite{SSphi}, 
can be obtained from the known form of the bosonic world-sheet Green 
function in string theory~\cite{Martinec}. Unlike in string theory, 
no universal expression for the Green function 
at any higher loop order is known in particle theory. 
After all, many topologically inequivalent $\Phi^3$ vacuum diagrams 
exist at high loop order, and likewise there are many topologically 
inequivalent ways of inserting two external legs on a given vacuum 
diagram. But for any insertion of two external legs in a specific 
vacuum diagram, corresponding to any given 2-point $(m+1)$-loop 
$\Phi^3$ diagram, we should be able to reproduce the particle 
Green function from the string Green function by approaching the 
singular point at the boundary of string moduli space where the 
string world-sheet degenerates into the desired $\Phi^3$ 
particle diagram.

We do this by a two-step procedure: To approach the singular point, 
where the surface degenerates into a world-line diagram, we first 
have to identify the neighbourhood of the singular point; i.e. the 
corner of moduli space where the string world sheet,which is a 
two-punctured genus $m+1$ Riemann surface, looks ``almost'' like 
the desired $\Phi^3$ particle diagram, i.e. consists of cylinders, 
very long in units of the diameter, that are joined together at the 
vertices. Such a corner of moduli space is defined by letting the 
$3m+2$ complex string moduli approach an appropriate limit, known as 
a {\em pinching limit}.
In the corner of moduli space defined by the pinching limit, the 
mapping of string moduli into Schwinger Proper Times (SPTs) of the 
particle diagram can be identified. Since SPTs have dimension of 
$(\mbox{length})^2$, they are all proportional to $\alpha'$. The 
formal scaling limit, leading to the particle diagram, is then 
obtained by taking $\alpha' \ra 0$, keeping all SPTs fixed at finite 
values. We refer to this as the {\em field theory limit} pertaining 
to the given $\Phi^3$ diagram.

The procedure for obtaining the particle Green function is this: 
Evaluate the string Green function $G^{(m)}_{\rm str} (z_1,z_2)$ 
in the pinching limit pertaining to the $\Phi^3$-diagram in 
question. By taking the formal field theory limit $\alpha' \ra 0$, 
keeping all SPTs fixed, we obtain the particle Green function, 
$\tilde{G}_B^{(m)} (\tau_1,\tau_2)$, as follows
\beq
G^{(m)}_{\rm str} (z_1,z_2) \ \alimit \ 
\frac{1}{\alpha'} \tilde{G}_B^{(m)} 
(\tau_1,\tau_2) + \ \mbox{finite} \ . \label{limit}
\eeq
However, there is an important qualification to this statement: As is 
well known, the bosonic Green function in string theory is not a 
scalar function of the two arguments; instead it depends on which 
choice of local coordinates we make on the Riemann surface in the 
vicinity of the two punctures where the operators are inserted. For 
this reason, we will not obtain the particle Green function from the 
string Green function in the field theory limit, as in \eq{limit}, 
unless we make a {\em proper choice of local coordinate} in the 
vicinity of the punctures for the degenerate world-sheet 
representing the $\Phi^3$ particle diagram in question. 

In this paper we present a general prescription on how to choose 
the local coordinates for any degenerate world-sheetresembling a 
$\Phi^3$ diagram, so as to obtain the desired relation \eq{limit}. 
We then proceed to verify eq.~\eq{limit} and our prescription, 
in particular, in some explicit examples.  
In their paper~\cite{SSphi}, Schmidt and Schubert found a very 
elegant form for the particle Green function pertaining to a large 
class of diagrams, namely any diagram obtained by inserting $2m+2$ 
external legs on a given circle and then joining together $m$ pairs 
of these legs by the insertion of an internal propagator to form a 
total of $m+1$ loops. As we choose different orderings of the $2m+2$ 
external legs inserted on the circle, many topologically different 
2-point $(m+1)$-loop diagrams are obtained. We show how the string 
Green function can be reduced to the Schmidt-Schubert particle 
Green function for this whole class of diagrams.

Other particle Green functions can be obtained if one (or both) of 
the external legs is inserted on one (or two) of the internal 
propagators instead of on the circle. Generalizing the derivation 
of Schmidt and Schubert, we also find the particle Green functions 
corresponding to these classes of diagrams, and we show that they, 
too, are obtained from the universal expression of the string Green 
function by taking the relevant field theory limit and using our 
prescription for the choice of local coordinates. 

Thus, we analyze all possible insertions in any $\Phi^3$ vacuum 
diagram that can be obtained by joining together $m$ pairs of legs 
inserted on a circle. (We shall refer to this as a Schmidt-Schubert 
type vacuum diagram). For $m \leq 4$ this exhausts all one-particle 
irreducible possibilities. For $m \geq 5$ more complicated vacuum 
diagrams can be imagined, but we do not attempt an analysis of these 
diagrams.

The paper is organized as follows: In section 2 we present the bosonic 
Green function in string theory and show how to identify the pinching 
limit corresponding to any given $\Phi^3$-diagram, using the Schottky 
parametrization~\cite{paoloschottky} of Riemann surfaces. We also 
define the mapping of moduli into SPTs and present our prescription 
for choosing the local coordinates around the punctures. In section 3 
we consider the various particle Green functions that describe all 
possible insertions in a $\Phi^3$ vacuum diagram of the 
Schmidt-Schubert type, and in section 4 we recover these Green 
functions from the string Green function through the relevant field 
theory limits. Finally, section 5 contains our conclusions, and we 
also comment on some of the problems that remain to be solved before 
a string-based set of Bern-Kosower like rules can 
be obtained for multiloop Yang-Mills theory.

\section{The Green Function in String Theory}
\setcounter{equation}{0}
\indent

The bosonic Green function in closed string theory at genus $m+1$ 
can be defined as the two point function
\beq
G^{(m)}_{\rm str} (P_1,P_2) = \langle X(w_1=0) X(w_2=0)\rangle\ , 
\label{Xdef}
\eeq
evaluated on a given world sheet described by $3m$ complex modular 
parameters.~\footnote{The ket-vacuum in \eq{Xdef} is the conformal 
one (i.e. a zero-momentum eigenstate), while the bra-vacuum is the 
state dual to this, i.e. a position eigenstate located at the 
origin.} 
Here $P_1$ and $P_2$ are two points on the world-sheet. Since the 
field $X$, defining the embedding of the world-sheet into target 
space, is not a genuine conformal field of dimension zero, the 
Green function defined by eq.~\eq{Xdef} depends on which local 
holomorphic coordinate system $w_i$ we use at the point $P_i$, 
$i=1,2$.

We may introduce a single holomorphic coordinate $z$, given in 
terms of the local coordinate $w_i$ by 
\beq
w_i = V_i^{-1}(z) \ \ \mbox{or} \ \ z = V_i (w_i) \ .
\eeq
It is conventional to take $w_i(P_i)=0$. This implies 
that $z_i \equiv z(P_i) = V_i(0)$, $i=1,2$.

In terms of the coordinate $z$ the Green function 
\eq{Xdef} may be written on the form~\cite{Martinec} 
\beq
G^{(m)}_{\rm str}(z_1,z_2)=\ln\left\vert
    {E(z_1,z_2) \over (V_1'(0) V_2'(0))^{1/2}} \right\vert  
-{1\over2}\sum_{\mu,\nu=0}^m\Omega_\mu\imtau{\mu\nu}\Omega_\nu\ , 
\label{no8} 
\eeq 
where $E(z_1,z_2)$ is the prime form, $\tau$ is the 
period matrix and
\beq
\Omega_{\mu} = {\mbox Re} \int_{z_1}^{z_2} \omega_{\mu} 
\ \ , \ \ 
\mu=0,1,\ldots,m \ , 
\eeq
is the real part of the $\mu$th abelian integral. (Our conventions 
are such that in a canonical homology basis 
$\{ a_{\mu},b_{\mu} \vert\mu 
= 0,1,\ldots,m\}$ we have $\oint_{a_{\mu}} 
\omega_{\nu} = 
2\pi i \delta_{\mu \nu}$ and $\oint_{b_{\mu}} 
\omega_{\nu} = 2\pi i 
\tau_{\mu \nu}$.) 

The dependence on the choice of local coordinates $w_1$ and $w_2$ 
enters through the quantities $V_i'(0)$ which appear because the 
prime form carries conformal dimension $-1/2$ with respect to both 
arguments. When the Green function is used to compute a physical 
string amplitude, the dependence on the local coordinates always 
drops out by virtue of the on-shell conditions for the external 
string states; but in the Green function {\em per se}, the dependence 
does not drop out. Therefore, before we can compare the string 
Green function \eq{no8}, evaluated in a corner of moduli space 
corresponding to a given $\Phi^3$ particle diagram, with the particle 
Green function pertaining to that diagram, we 
have to specify our choice of local coordinates.
\subsection{Choosing the Local Coordinates}
\indent

In general, the local coordinate $w_i$ may be chosen to depend on 
all the moduli of the world-sheet, and even on the positions of 
the vertex operators representing external states. However, only a 
proper choice of local coordinate will give rise to a string Green 
function that reproduces the particle Green function, as in 
\eq{limit}, in the corners of moduli space
corresponding to $\Phi^3$ particle diagrams. Our task is 
to find out which is the proper choice.

To this end it is useful to first consider the simplest example of 
all, where the world-sheet on which the two $X$-operators are 
inserted is just an infinite cylinder, representing a freely 
propagating string. In this case the particle diagram is an 
infinite line, parametrized by a Schwinger proper time $\tau$, 
and the particle Green function is
\beq
G_B^{\rm line} (\tau_1,\tau_2) = \vert \tau_1 - \tau_2 \vert  \ , 
\label{free}
\eeq
as can be seen for example by taking the limit $T \ra \infty$ in the 
known expression for the particle Green function of a loop of total
SPT length $T$, given by \cite{Str}, \cite{poly}
\beq
G_B (\tau_1,\tau_2) = \vert \tau_1 - \tau_2 \vert - 
\frac{(\tau_1-\tau_2)^2}{T} \ . \label{no3}
\eeq
In string theory, the Green function \eq{no8} reduces in 
the case of zero loops to
\beq
G_{\rm str}^{\rm tree} (z_1,z_2) =  
\ln\left\vert
    {z_1-z_2 \over (V_1'(0) V_2'(0))^{1/2}} \right\vert \ .
\label{stringgf}
\eeq
In order to relate the two Green functions \eq{stringgf} 
and \eq{free} we first have to establish the relation between the 
coordinates $z_i$ and the SPTs $\tau_i$.

In the customary parametrization of the cylinder, where the infinite 
future corresponds to $z=\infty$ and the infinite past to $z=0$,
we have $z=\exp\{t + i \sigma\}$, where $t$ is the time-coordinate 
on the world-sheet and $\sigma \in [0;2\pi]$ labels the points on 
the closed string at fixed time. The relation between SPT $\tau$ and 
the coordinate $t$ can be found by requiring that the operator 
generating the time translation $t \ra t' = t - \delta t$, 
which is given by $\delta t (L_0 + \bar{L}_0)$, assumes the standard 
form $\delta \tau (p^2 + m^2)$ when acting on any string state 
satisfying the level-matching condition 
$L_0 = \bar{L}_0 = \frac{\alpha'}{4} (p^2 + m^2)$. One finds
\beq
\tau = \frac{\alpha'}{2} t + \tau_0  = \frac{\alpha'}{2} 
\ln \vert z \vert + \tau_0 \ , \label{SPTrel}
\eeq
where $\tau_0$ is an unspecified additive constant.

In a more general parametrization, where the infinite future and 
infinite past are represented by $z=z_{\alpha}$ and 
$z=z_{\beta}$ respectively, the relation \eq{SPTrel} becomes
\beq
\tau = \frac{\alpha'}{2} \ln \left\vert 
\frac{z-z_{\beta}}{z-z_{\alpha}} \right\vert 
+ \mbox{constant} = \frac{\alpha'}{2} \mbox{Re} \int^z \omega \ ,
\label{SPTrel2}
\eeq
where
\beq
\omega(z) = \frac{1}{z-z_{\beta}} - \frac{1}{z-z_{\alpha}} = 
\frac{z_{\beta}-z_{\alpha}}{(z-z_{\alpha})(z-z_{\beta})} 
\label{oneform}
\eeq
is the {\em unique} holomorphic one-form on the cylinder 
satisfying the normalization condition
\beq
\int_a \omega = 2\pi i \ , \label{normaliz} 
\eeq
$a$ being the homology cycle encircling the cylinder 
(represented in the $z$-plane by any curve encircling 
$z_{\beta}$ but not $z_{\alpha}$).

We now consider the proper choice of local coordinate. 
By definition
\beq
(V_i'(0))^{-1} = \left. \frac{{\rm d} w_i}{{\rm d} z} 
\right|_{z=z_i}
\eeq
carries conformal dimension one with respect to $z_i$. 
This observation makes it very tempting to identify 
$(V_i'(0))^{-1}$ with the unique holomorphic one-form $\omega$, 
evaluated at the point $z_i$, i.e. to take
\beq
V_i'(0) =  (\omega(z_i))^{-1} = 
\frac{(z_i-z_{\alpha})(z_i-z_{\beta})}{z_{\beta}-z_{\alpha}}  \ . 
\label{choice}
\eeq
As we now proceed to show, this choice does indeed ensure that 
the string Green function \eq{stringgf} reduces to the particle 
Green function \eq{free} in the field theory limit, as in \eq{limit}.
Choosing for simplicity the constant in eq.~\eq{SPTrel2} 
to be zero, it follows that in the limit where $\alpha' \ra 0$ 
for fixed $\tau$, the point $z$ must approach either $z_{\alpha}$ or 
$z_{\beta}$ depending on whether $\tau$ is positive or 
negative. Since we are free to choose the location of $\tau=0$, we 
may assume $\tau_1$ and $\tau_2$ to be positive without loss of 
generality, so that $z_1$ and $z_2$ both approach $z_{\alpha}$. 
In this limit
\beq
\tau_1 - \tau_2 = 
\frac{\alpha'}{2} \ln \vert 
\frac{z_1-z_{\beta}}{z_1-z_{\alpha}}
\frac{z_2-z_{\alpha}}{z_2-z_{\beta}} \vert \simeq
- \frac{\alpha'}{2} \ln \vert 
\frac{z_1-z_{\alpha}}{z_2-z_{\alpha}} \vert \ 
, \label{no15b}
\eeq
and we see that there are two orderings to consider, namely 
$\tau_1 > \tau_2$, corresponding to 
$ |z_1-z_{\alpha}| \ll |z_2-z_{\alpha}|$, and
$\tau_2 > \tau_1$, corresponding to  
$ |z_2-z_{\alpha}| \ll |z_1-z_{\alpha}|$. 
In the first case we find
\beq
G_{\rm str}^{\rm tree} (z_1,z_2) \ \alimit \  
\ln\left\vert \frac{z_2-z_{\alpha}}{z_1-z_{\alpha}} 
\right\vert^{1/2} = \frac{1}{\alpha'} (\tau_1-\tau_2) \ , 
\eeq
and in the second case we obtain the same result, with labels $1$ 
and $2$ interchanged. In summary, we find the desired behaviour
\beq
G_{\rm str}^{\rm tree} (z_1,z_2) \ \alimit \  
\frac{1}{\alpha'} \vert \tau_1 - \tau_2 \vert \ = \
\frac{1}{\alpha'} G_B^{\rm line} (\tau_1,\tau_2) \ ,
\eeq
regardless of the ordering chosen for the SPTs.

Our task is to generalize the prescription \eq{choice} to any 
world-sheet resembling a particle $\Phi^3$ diagram. But how can 
$\omega$ be specified on a world-sheet containing several external 
legs and handles, where the holomorphic one-forms span a vector space 
of complex dimension greater than one?
Fortunately, for the degenerate world-sheets resembling $\Phi^3$ 
vacuum diagrams there is an essentially unique way to do this:
Any vacuum $\Phi^3$ diagram consists of propagators joined together 
at 3-point vertices. Any propagator having a fixed SPT length 
corresponds to a cylinder that becomes infinitely long (in units of 
the diameter) in the limit $\alpha' \ra 0$. As we have seen, there 
is a {\em unique} one-form $\omega$ that is holomorphic at all points 
of an infintely long cylinder and satisfies the normalization 
condition \eq{normaliz}. Therefore, our prescription for choosing 
$V_i'(0)$ for a world-sheet resembling a given $\Phi^3$-diagram, with 
the point $z_i$ sitting on some very long cylinder corresponding to 
an internal propagator of the vacuum diagram, is simply to take
\beq
V_i'(0) = (\omega(z_i))^{-1} \ , \label{omegachoice}
\eeq
where $\omega$ is {\em any} one-form that is: i) holomorphic at 
all points of the cylinder on which the point $z_i$ is located; 
and ii) satisfies the normalization condition \eq{normaliz}, 
$a$ being a closed contour encircling the cylinder. In the limit 
$\alpha' \ra 0$, where the cylinder becomes infinitely long, all 
such choices will be equivalent.

\subsection{Identifying the Proper Pinching Limit}
\indent

In this subsection we show how to identify a pinching limit
(i.e. a corner of moduli space) where the string world-sheet 
degenerates into any given N-point $(m+1)$-loop $\Phi^3$ particle 
diagram. Of course, for the analysis of the Green function we 
only need the case $N=2$, but since the generalization to arbitrary 
$N$ does not present any further complications, we may as 
well be general.

As explained in more detail in refs.~\cite{paperone},\cite{me2}, 
the relevant pinching limit can be identified by carefully 
constructing the degenerate world-sheet using the sewing procedure 
of refs.~\cite{Martinec},\cite{Jenslyng}. In this paper we formulate 
a recipe which allows one to identify the pinching limit 
corresponding to a given diagram without having to go into the 
subtleties of the sewing procedure.

To be specific, we need a concrete parametrization of moduli space. 
The Schottky parametrization~\cite{paoloschottky} 
of Riemann surfaces turns out to be very useful for our purposes. 
In this parametrization, the world-sheet of the string is described 
as a sphere with $m+1$ pairs of holes, and each pair of holes 
is connected by a tube, thus forming a handle. The positions of the
two holes forming the $\mu$th pair  is parametrized by points 
$z_{\alpha_{\mu}}$ and $z_{\beta_{\mu}}$ in the compactified 
complex plane ${\bf C} \cup \{ \infty \}$, and the length
of the closed loop that we form by the insertion of the
connecting tube is described by the modulus of a complex 
parameter $k_{\mu}$, known as a {\em multiplier}.

The relation between the multiplier $k_{\mu}$ and the {\em total} 
Schwinger proper time $T_{\mu}$ of the $\mu$th closed loop, formed 
by joining together the $\mu$th pair of holes, is given by
\beq 
T_{\mu} = - \frac{\alpha'}{2} \ln \vert k_{\mu} \vert \ , 
\label{deltai}
\eeq
and in the particle limit, where we take $\alpha' \ra 0$ while 
keeping $T_{\mu}$ fixed, all multipliers become infinitely small
\beq
     k_{\mu} \rightarrow 0 \ \ , \ \ \mu=0,1,\ldots,m \ .
\eeq
When all the multipliers are approaching zero, the various 
quantities defining the Green function 
\eq{no8} assume the following simple form
\begin{eqnarray}
E(z_1,z_2) & = & z_1 - z_2 + {\cal O} (k_{\mu}) \ , 
\label{no8b} \\
(2\pi\mbox{Im}\tau)_{\mu\nu} & = & 
-\delta_{\mu\nu}\ln\vert{k_\mu}\vert 
-(1-\delta_{\mu\nu})\ln  \left\vert
{z_{\alpha_\mu}-z_{\alpha_\nu}\over 
z_{\alpha_\mu}-z_{\beta_\nu}}
{z_{\beta_\mu}-z_{\beta_\nu}\over 
z_{\beta_\mu}-z_{\alpha_\nu}}  
                          \right\vert + {\cal O} 
(k_{\mu},\bar{k}_{\mu}) \ , 
\label{no9} \\
\Omega_\mu  & = & \ln \left\vert
{z_2-z_{\beta_\mu}\over z_1-z_{\beta_\mu}}
{z_1-z_{\alpha_\mu}\over z_2-z_{\alpha_\mu}} \right\vert 
+ {\cal O} 
(k_{\mu},\bar{k}_{\mu}) \ . \label{no10}
\end{eqnarray}
\indent

However, to fully define the pinching limit corresponding to a 
given $N$-point $(m+1)$-loop $\Phi^3$ particle diagram, just 
taking the multipliers to be very small does not suffice.
We still have to specify the behaviour of the $2m+N+2$ points 
$z_i$, $i=1,\ldots,N$ and $\zalmu,\zbemu$, $\mu=0,1,\ldots,m$. 
This we do as follows:
\begin{figure}
\begin{center}
\input{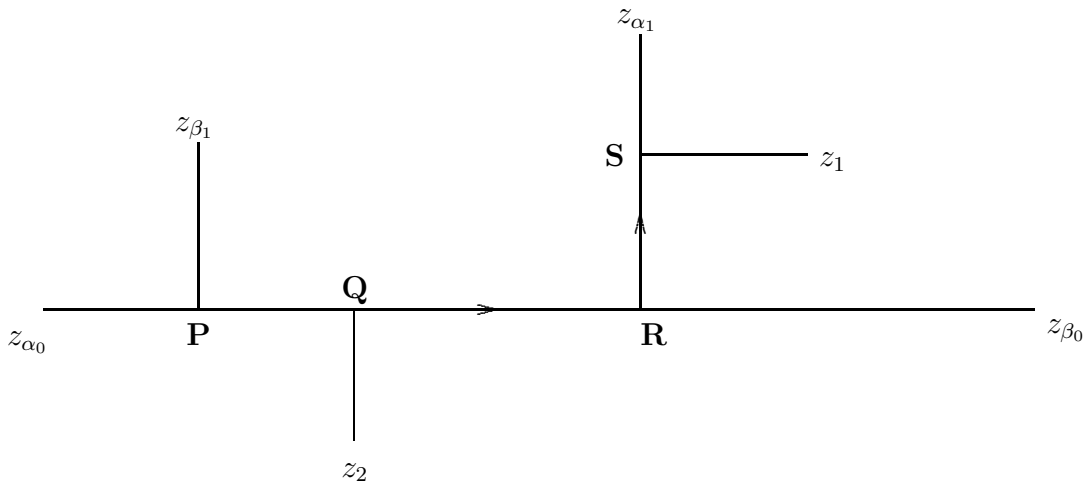}
\end{center}
\caption[gyh]{Example of a tree with one side-branch. 
The branch endpoints are $z_{B_0}=z_{\beta_0}$ and 
$z_{B_1}=z_{\alpha_1}$. The vertices $P,Q,R$ and $S$ are labelled 
by $z_{\beta_1}, z_2, z_{\alpha_1}$ and $z_1$ respectively.}
\end{figure}
First we cut open all the loops of the diagram so as to form a 
$2m+2+N$-point $\Phi^3$ tree diagram, ${\cal T}$, where the 
original external legs are labelled by $z_i$, $i = 1,\ldots,N$ 
and the pair of external legs formed by cutting the $\mu$th loop 
by $\zalmu$ and $\zbemu$, $\mu=0,1,\ldots,m$ 
(for example, see Fig.1). Projective invariance allows us to fix, 
say, $z_{\alpha_0} = \infty$ and 
$z_{\beta_0} = 0$.~\footnote{We could actually fix the position of 
a third point, but we prefer not to do so.} 
Then the remaining $2m+N$ external legs, which we call the 
{\em movable external legs}, are labelled by the points $z_i$, 
$i=1,\ldots,N$ and $z_{\alpha_j},z_{\beta_j}$, $j=1,\ldots,m$, and 
we want to specify how to pinch these points. To this end it is 
very helpful to think of some flow of SPT, $\tau$, starting at the 
leg labelled by $z_{\alpha_0} = \infty$ 
and flowing towards the future. We follow the flow all 
the way to the future endpoint which we take to be at the leg 
labelled by $z_{\beta_0}=0$. This defines what we call the 
{\em main branch}, $B_0$, of the tree diagram ${\cal T}$. 
The endpoint of the branch $B_0$ we also denote by $z_{B_0}$, i.e. 
$z_{B_0} = z_{\beta_0}$. In general not all the remaining external 
legs will be inserted as stems on the main branch; instead the 
diagram will have one or more {\em side-branches} 
$B_1, B_2, \ldots $. Each side-branch $B_i$ has its own flow of SPT, 
$\tau^{(i)}$, which we take to begin (with the value $\tau^{(i)}=0$)
at the vertex where the side-branch is joined to the main branch. 
Now, the side-branch $B_i$ ends with {\em two} external legs, and 
so we choose one of them to represent the endpoint, $z_{B_i}$, of 
the side-branch $B_i$, and of the SPT flow $\tau^{(i)}$, and the 
other is then considered to be a stem inserted on the side-branch.

In exactly the same way, the side-branches may themselves have 
side-branches, which give rise to new SPT flows. At the end we have 
decomposed the entire tree diagram into the main branch $B_0$ and a 
set of side-branches $B_i$, $i=1,\ldots,N_b$, each having its own
SPT flow. The direction of the SPT flow $\tau^{(i)}$ 
(towards the endpoint $z_{B_i}$) defines a SPT ordering of the 
vertices along the branch $B_i$. 

Furthermore, the $2m+N$ movable external legs have been divided into 
side-branch endpoints $z_{B_i}$, $i=1,\ldots,N_b$, and stems inserted 
on the branches. Keeping this in mind we may now establish a one-to-one 
correspondence between the vertices of the diagram and the points 
$z_1,\ldots,z_N;z_{\alpha_1},z_{\beta_1},
\ldots,z_{\alpha_m},z_{\beta_m}$ 
labelling the movable external legs. Simply, if 
$z \in \{z_1,\ldots,z_N;z_{\alpha_1},z_{\beta_1},
\ldots,z_{\alpha_m},z_{\beta_m} \}$ 
labels a stem inserted on a branch, then $z$ is associated with the 
corresponding vertex. Instead, if $z$ labels a side-branch endpoint, 
then $z$ is associated with the vertex at which the side-branch is 
attached. An example is shown in Fig.1.

The pinching limit corresponding to the given diagram is now 
identified as follows: The $z$-values associated to the vertices of 
the branch $B_i$, $i=0,1,\ldots,N_b$, all approach the branch 
endpoint $z_{B_i}$, i.e. 
\beq
       \vert z - z_{B_i} \vert \ll 1 \ . \label{zzbi}
\eeq
Furthermore,  $z$-values associated to vertices having ``late'' 
values of $\tau^{(i)}$ approach $z_{B_i}$ much faster than 
$z$-values associated to vertices having ``early'' values of 
$\tau^{(i)}$. Thus, if $z$ and $\tilde{z}$ are associated with two 
vertices along the branch $B_i$, having SPTs $\tau^{(i)}$ and 
$\tilde{\tau}^{(i)}$ respectively, then
\begin{eqnarray}
\vert z - z_{B_i} \vert \ll \vert \tilde{z} - z_{B_i} \vert \ll 1 & 
\mbox{if} & \tau^{(i)} > \tilde{\tau}^{(i)} \label{pinch} \\
\vert \tilde{z} - z_{B_i} \vert \ll \vert z - z_{B_i} \vert \ll 1 & 
\mbox{if} & \tilde{\tau}^{(i)} > \tau^{(i)} \ . \nonumber
\end{eqnarray}
One should note that in general there will be many different pinching 
limits that give rise to the same $\Phi^3$-diagram. For example, the 
pinching limit that we obtain by interchanging $\zalmu$ and $\zbemu$ 
will rather obviously define the same diagram 
-- after all, when we form two external legs by cutting open the 
$\mu$th loop, it is a matter of pure convention which one is labelled 
by $\zalmu$ and which one by $\zbemu$. Similarly, 
any pinching limit that is obtained by permuting the $m$ triplets 
$\{k_j,z_{\alpha_j},z_{\beta_j}\}$ will also 
correspond to the same $\Phi^3$-diagram, since one is 
just changing the numbering of the loops.

The existence of several equivalent pinching limits merely reflects 
the fact that we are working in Teichm\"{u}ller space: The equivalent 
pinching limits are related by modular transformations and correspond 
to the same corner of moduli space. One is therefore free to choose 
any one of the equivalent pinching limits to represent the corner of 
moduli space corresponding to a given $\Phi^3$-diagram.

In section 4 we provide some explicit examples that may help to 
clarify the general procedure for identifying a pinching limit that 
we have given in this section.

\subsection{The Mapping of String Moduli into SPT}
\indent

In the previous subsection we already introduced a SPT 
parametrization of any given $\Phi^3$ particle diagram: Along the 
main branch we use the SPT $\tau\equiv \tau^{(0)}$, flowing from 
$z_{\alpha_0}$ towards $z_{\beta_0}$, and along the side-branch 
$B_i$ we define the SPT $\tau^{(i)}$, starting at zero value at 
the vertex where the side-branch is attached to the main part of 
the tree and flowing towards the endpoint $z_{B_i}$. We also 
identified a pinching limit describing the corner in string moduli 
space where the world-sheet degenerates into the particle diagram.

In this subsection we present the precise mapping between the points 
of the set $M \equiv \{z_1,\ldots,z_N;z_{\alpha_1},z_{\beta_1},
\ldots,z_{\alpha_m},z_{\beta_m} \}$ 
and the SPTs of the vertices in the particle diagram. Remember that 
to each vertex we have associated a point  $z \in M$. Then, if two 
vertices belonging to the branch $B_i$ have SPTs $\tau^{(i)}$ and 
$\tilde{\tau}^{(i)}$ and are associated to the points $z$ and 
$\tilde{z}$ respectively, the relation, like in eq.~\eq{no15b}, is 
simply given by
\beq
\tau^{(i)} - \tilde{\tau}^{(i)} = -\frac{\alpha'}{2}\ln \left\vert 
\frac{z-z_{B_i}}{\tilde{z}-z_{B_i}} \right\vert \ . 
\label{SPTdef}
\eeq
It is clear that with this precise identification, the pinching limit 
\eq{pinch} amounts to the statement that $\vert \tau^{(i)} - 
\tilde{\tau}^{(i)} \vert / \alpha' \gg 1$, i.e. that the cylinder 
connecting the two vertices is very long in string units. As we 
recall, the formal field theory limit is obtained by taking 
$\alpha' \ra 0$ for fixed SPT values.

The expression \eq{SPTdef} only defines {\em differences} of SPT 
values along the branch $B_i$. It does not provide the relation 
between the points of the set $M$ and the SPT of a single vertex.
\begin{figure}
\begin{center}
\input{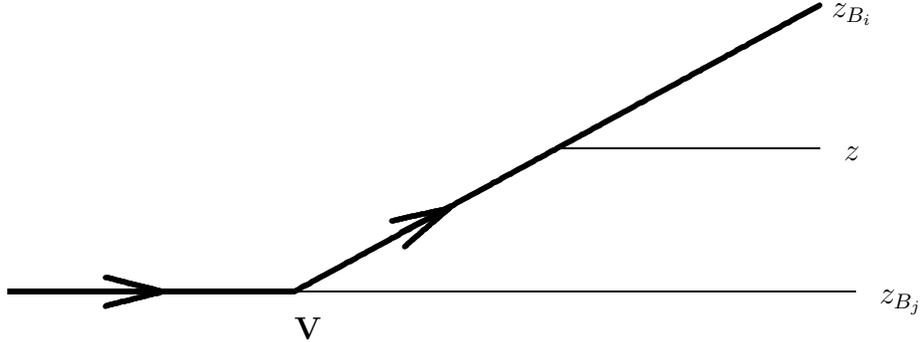}
\end{center}
\caption[gyhh]{How to obtain eq.~\eq{SPTabs}. The thick line 
represents the new SPT flow. See the text for details.}
\end{figure}
For vertices along the main branch $B_0$, we simply take
\beq
\tau = -\frac{\alpha'}{2} \ln \vert z \vert \ ,  
\label{mainspt}
\eeq
while for vertices along the side-branch $B_i$, 
$i=1,\ldots,N_b$, the relation is
\beq
\tau^{(i)} = - \frac{\alpha'}{2} \ln \left\vert 
\frac{z-z_{B_i}}{z_{B_j}-z_{B_i}} \right\vert \ , 
\label{SPTabs}
\eeq
where $B_j$, $j\in \{0,1,\ldots,i-1,i+1,\ldots,N_b\}$, is the 
branch to which $B_i$ is attached. To derive eq.~\eq{SPTabs} we may 
refer back to the sewing procedure~\cite{Martinec},\cite{Jenslyng}, 
but the following argument is easier (see Fig.2): We consider 
the side-branch $B_i$ to be attached to the branch $B_j$ at the 
vertex $V$, and according to this decomposition of the tree into 
branches, the SPT flow $\tau^{(i)}$ starts at the vertex $V$. Now, 
the vertex $V$ divides the branch $B_j$ into two: speaking in terms 
of the SPT flow  $\tau^{(j)}$ these are the parts of the branch 
$B_j$ happening ``before'' and ``after'' the vertex. This division 
allows us to define a new decomposition of the tree diagram into 
branches, one where we consider the part of the old branch $B_j$ 
happening {\em after} the vertex $V$ to be a side-branch 
inserted on the branch formed by combining the part of the old 
branch $B_j$ happening {\em before} the vertex with the old 
side-branch $B_i$. Corresponding to this new decomposition of the 
diagram into branches, we have a new SPT, $\tau^{(i)}_{\rm new}$, 
which is zero at the point of the diagram where the old SPT 
$\tau^{(j)}$ was zero, but which still runs towards the point 
$z_{B_i}$. Furthermore, with the new decomposition, both the vertex 
$V$, which according to our rules is now associated with the point 
$z_{B_j}$, as well as the vertex associated with the point $z$, 
now belong to the branch $B_i$, and therefore we may use 
eq.~\eq{SPTdef} to compute the SPT interval 
$\Delta \tau^{(i)}_{\rm new}$
between the two vertices. Since by definition this interval is 
nothing but the value taken by the old SPT $\tau^{(i)}$ at the 
vertex associated with $z$, we arrive at the expression \eq{SPTabs}.

Since $\tau^{(i)}$ is positive and very large in units of $\alpha'$,
the relation \eq{SPTabs} implies that the pinching limit \eq{zzbi} 
is further specified by the inequality
\beq
\vert z - z_{B_i} \vert \ll \vert z_{B_j} - z_{B_i} \vert \ .
\eeq
\section{The Green Function in Particle Theory}
\setcounter{equation}{0}
\indent

Schmidt and Schubert have obtained the bosonic worldline Green 
function for the class of multi-loop diagrams where all vertices 
are located on a single fundamental loop. Their Green function was 
originally derived in the scalar $\Phi^3$ theory \cite{SSphi} and 
was subsequently applied to a specific type of multiloop 
photon diagrams in spinor QED \cite{SSqed}. In the worldline 
formulation of $\Phi^3$ theory, the sum of all $N$-point $(m+1)$-loop 
Feynman diagrams that have their vertices located on a single loop 
is expressed by the following integral over SPTs
(omitting for simplicity the coupling constant and mass factors):
\begin{eqnarray}
\Gamma_N^{(m+1)}& = & \int_0^{\infty}{dT\over T}T^{-D/2} 
(4\pi)^{-{D\over2}(m+1)}
\prod_{i=1}^m \int_0^{\infty}d{\bar T}_i
\int_0^T d\tau_{\alpha_i} \int_0^T d\tau_{\beta_i} 
\prod_{n=1}^N \int_0^T d\tau_n \nonumber \\
& & \times \ (\mbox{det}A)^{-D/2} \ 
\exp[{1\over2}\sum_{k,l=1}^N
          p_k p_l G^{(m)}_B(\tau_k,\tau_l)]      \ .     
    \label{no1}
\end{eqnarray}  
Here all points on the fundamental loop are labelled by a SPT 
$\tau \in [0;T]$. Thus, 
$\tau_1,\ldots,\tau_N$ are the points where the external legs have
been inserted and the extra loops have been obtained by joining 
together the points $\tau=\tau_{\alpha_i}$ and $\tau=\tau_{\beta_i}$ 
by means of an internal propagator of SPT length $\bar{T}_i$, 
$i=1,\ldots,m$. The worldline Green function is given by
\begin{eqnarray}
\lefteqn{G^{(m)}_B(\tau_1,\tau_2) = } \label{no2} \\
& & G_B(\tau_1,\tau_2)+{1\over2}\sum_{i,j=1}^m  
  [G_B(\tau_1,\tau_{\alpha_i})-G_B(\tau_1,\tau_{\beta_i})]
A^{-1}_{ij}
 [G_B(\tau_2,\tau_{\alpha_j})-G_B(\tau_2,\tau_{\beta_j})] \ ,  
\nonumber
\end{eqnarray}
where $A$ is the symmetric $m\times m$ matrix defined 
by~\footnote{Notice that
our matrix $A$ is the inverse of the matrix $A$ used in 
refs.~\cite{SSphi},\cite{SSqed}.}
\beq
 A_{ij}\equiv{\bar T}_i\delta_{ij} -
         X(\tau_{\alpha_i},\tau_{\beta_i};
           \tau_{\alpha_j},\tau_{\beta_j}) \ ,           
    \label{no4}
\eeq
\beq
X(\tau_a,\tau_b;\tau_c,\tau_d)\equiv{1\over2}
        \left[ G_B(\tau_a,\tau_c)-G_B(\tau_a,\tau_d)
       -G_B(\tau_b,\tau_c)+G_B(\tau_b,\tau_d)\right] \ . 
    \label{no5} 
\eeq                             
In these formulae $G_B$ is the one-loop Green function 
given by eq.~\eq{no3}.

Actually, what we obtain from the string Green function by taking 
the field theory limit is not $G_B^{(m)}$, but rather the
redefined Green function 
\beq
{\tilde 
G}^{(m)}_B(\tau_1,\tau_2)=G^{(m)}_B(\tau_1,\tau_2)
               -{1\over2}G^{(m)}_B(\tau_1,\tau_1)
               -{1\over2}G^{(m)}_B(\tau_2,\tau_2) \ ,    
     \label{no6}
\eeq
which by construction satisfies ${\tilde G}^{(m)}_B(\tau,\tau)=0$.
In eq.~\eq{no1} we can replace $G_B^{(m)}$ by $\tilde{G}_B^{(m)}$, 
since the extra terms in the exponent cancel due to conservation of 
the external momenta. By using eq.~\eq{no2} we find
\beq
{\tilde G}^{(m)}_B(\tau_1,\tau_2)=G_B(\tau_1,\tau_2)
-\sum_{i,j=1}^m X(\tau_1,\tau_2; 
\tau_{\alpha_i},\tau_{\beta_i})A^{-1}_{ij}
X(\tau_1,\tau_2; \tau_{\alpha_j},\tau_{\beta_j}) \ .     
     \label{no7}
\eeq

Other types of Green functions appear if we want to consider the 
situation where external legs are inserted not only on the 
fundamental loop but also on the internal propagators. These Green 
functions can be obtained by a straightforward generalization of 
the path-integral derivation leading to the expression \eq{no1}. 
This expression was obtained by Schmidt and Schubert by
computing an open path integral ${\cal D} x^{(i)}$, 
$i=1,\ldots,m$ for each internal propagator and finally 
a closed path integral ${\cal D} x$ for the fundamental loop, with 
the insertion of $N$ vertex operators 
(see ref.~\cite{SSphi} for details)
\beq
\int_0^T {\rm d} \tau_n \exp [ i p_n x(\tau_n) ] \ , \ \ 
\  n=1,\ldots,N \ ,  \label{vertexop}
\eeq
corresponding to the source 
\beq
J(\tau)=i\sum_{n=1}^N p_n\delta(\tau-\tau_n).
\eeq 
To obtain the other types of Green functions we imagine 
inserting vertex operators 
\beq
\int_0^{\bar{T}_j} {\rm d} \tau_n^{(j)} \exp [i p_n^{(j)} x^{(i)}
(\tau_n^{(j)})] \ , \ \ \ n=1,\ldots,N_j \ ,
\eeq
into the open path integral pertaining to the $j$'th internal 
propagator, $j=1,2,...,m$ (total number of external legs 
is now $N=N_L+\sum N_i$). This obviously modifies the result of
computing the open path integrals, giving rise to a more 
complicated source
\beq
J(\tau)=i\sum_{n=1}^{N_L}p_n\delta(\tau-\tau_n) + 
i\sum_{i=1}^m\sum_{n=1}^{N_i}p_n^{(i)}\left[ 
\delta(\tau-\tau_{\alpha_i})
+{\tau_n^{(i)}\over{\bar T}_i}\{ 
\delta(\tau-\tau_{\beta_i}) -
\delta(\tau-\tau_{\alpha_i}) \} \right],
\eeq
as well as an additional exponential factor $\exp(I)$, where
\beq
I={1\over2}\sum_{i=1}^m\sum_{n,n'=1}^{N_i}p_n^{(i)} 
p_{n'}^{(i)} 
\{ 
\vert\tau_n^{(i)}-\tau_{n'}^{(i)}\vert-\tau_n^{(i)}-\tau
_{n'}^{(i)}
 +{2\tau_n^{(i)}\tau_{n'}^{(i)}\over{\bar T}_i}\}.
\eeq
Finally, performing the path integral for the fundamental loop, 
one obtains the following exponential factor from which the 
Green functions can be read
\begin{eqnarray}
\lefteqn{\exp(I) \exp \left[ -{1\over2}\int_0^T d\tau d\tau' 
J(\tau)G_B^{(m)}(\tau,\tau')J(\tau') 
\right] \equiv } \label{exponent} \\
& & 
\exp\left[{1\over2}\sum_{i,j=0}^m\sum_{n=1}^{N_i}\sum_{n'=1}^{N_j}
p_n^{(i)} p_{n'}^{(j)} 
G_{ij}^{(m)}(\tau_n^{(i)},\tau_{n'}^{(j)})\right] \ . \nonumber
\end{eqnarray}
(Here we introduced the convenient and rather obvious notation 
$\tau_n^{(0)} \equiv \tau_n$ and $p_n^{(0)} \equiv p_n$, 
$n=1,\ldots,N_L\equiv N_0$ ).
After some rearrangement similar to \eq{no6}, using conservation of 
external momentum to obtain 
$G_{ii}^{(m)}(\tau^{(i)},\tau^{(i)}) = 0 $ for 
$i=0,1,\ldots,m$,  we find
\beq
G_{00}^{(m)}(\tau_1,\tau_2)  =  
                      {\tilde G}_B^{(m)}(\tau_1,\tau_2), 
\label{gm00} 
\eeq
\beq
G_{ii}^{(m)}(\tau_1^{(i)},\tau_2^{(i)})  =
\vert\tau_1^{(i)}-\tau_2^{(i)}\vert - 
{(\tau_1^{(i)}-\tau_{2}^{(i)})^2
\over{\bar T}_i} (1+{1\over{\bar T}_i}X^{(m)}_{ii}), 
\label{gmii}
\eeq
\beq
G_{i0}^{(m)}(\tau_1^{(i)},\tau_2)  =   
 \tau_1^{(i)} + {\tilde 
G}_B^{(m)}(\tau_{\alpha_i},\tau_2)  
- {{\tau_1^{(i)}}^2\over{\bar T}_i}(1+{1\over{\bar 
T}_i}X^{(m)}_{ii}) +2{\tau_1^{(i)}\over{\bar T}_i}
X^{(m)}(\tau_{\alpha_i},\tau_2;\tau_{\alpha_i},\tau_{\beta_i}),
\label{gmi0}
\eeq
\beqa
\lefteqn{G_{ij}^{(m)}(\tau_1^{(i)},\tau_2^{(j)})  =  
 \tau_1^{(i)} +
{1\over2}{\tilde G}_B^{(m)}(\tau_{\alpha_i},\tau_{\alpha_j}) 
- {{\tau_1^{(i)}}^2\over{\bar T}_i}
(1+{1\over{\bar T}_i}X^{(m)}_{ii}) }           \label{gmij} \\
&& + {\tau_1^{(i)}\tau_2^{(j)}\over{\bar T}_i{\bar T}_j}
X^{(m)}_{ij} + 2{\tau_1^{(i)}\over{\bar T}_i}
X^{(m)}(\tau_{\alpha_i},\tau_{\alpha_j};\tau_{\alpha_i},
\tau_{\beta_i}) 
+ (1,i\leftrightarrow 2,j \quad\mbox{for all terms}),\nonumber
\eeqa
where, in analogy with eq.~\eq{no5}, we have defined
\begin{eqnarray}
\lefteqn{X^{(m)} (\tau_a,\tau_b;\tau_c,\tau_d)  \equiv } 
\label{no46b} \\
& &  \frac{1}{2} \left[
G_B^{(m)} (\tau_a,\tau_c) - G_B^{(m)} (\tau_a,\tau_d) - 
G_B^{(m)} 
(\tau_b,\tau_c) + G_B^{(m)} (\tau_b,\tau_d) \right] \ , 
\nonumber
\end{eqnarray}
and we also introduced the notation
\beq
X^{(m)}_{ij}\equiv  
X^{(m)}(\tau_{\alpha_i},\tau_{\beta_i};
\tau_{\alpha_j},\tau_{\beta_j}) \ \ \ \mbox{and} \ \ \ 
X_{ij}\equiv  X(\tau_{\alpha_i},\tau_{\beta_i};
\tau_{\alpha_j},\tau_{\beta_j})\ .
\label{short1}
\eeq
Here the convention is that the SPT $\tau^{(i)}$ starts at zero 
value at the vertex $\tau=\tau_{\alpha_i}$ and ends (at value 
$\bar{T}_i$) at the vertex $\tau = \tau_{\beta_i}$.  Eqs.\eq{gmii} 
and \eq{gmi0} are obtained in ref.~\cite{sato} in the $m=1$ case.

In the Appendix we prove the following two identities:
\beq
A^{-1}_{ij}={1\over{\bar T}_i{\bar T}_j}
          ({\bar T}_i\delta_{ij}+X^{(m)}_{ij}), 
\label{app1}
\eeq
\beq
\sum_{k=1}^mA^{-1}_{ik}
X(\tau_a,\tau_b;\tau_{\alpha_k},\tau_{\beta_k})= 
{1\over{\bar T}_i}X^{(m)}(\tau_a,\tau_b;
\tau_{\alpha_i},\tau_{\beta_i}) \ , \label{app2}
\eeq
and by using these, together with eq.~\eq{no7}, we can rewrite 
the Green functions \eq{gmii} - \eq{gmij} on a more compact form:
\beq
G_{ii}^{(m)}(\tau_1^{(i)},\tau_2^{(i)})  =
\vert\tau_1^{(i)}-\tau_2^{(i)}\vert - 
(\tau_1^{(i)}-\tau_{2}^{(i)})^2 A^{-1}_{ii}, \label{agmii}
\eeq
\beqa
G_{i0}^{(m)}(\tau_1^{(i)},\tau_2) & = &  
\tau_1^{(i)}+ G_B(\tau_{\alpha_i},\tau_2)  \nn \\
& & - \sum_{k,l=1}^m 
[ -\tau_1^{(i)}\delta_{ik} +X_k(\tau_{\alpha_i},\tau_2)] 
A^{-1}_{kl}
[ -\tau_1^{(i)}\delta_{il} +X_l(\tau_{\alpha_i},\tau_2)]
,\label{agmi0}
\eeqa
\beqa
\lefteqn{G_{ij}^{(m)}(\tau_1^{(i)},\tau_2^{(j)})  =  
\tau_1^{(i)}+\tau_2^{(j)}+G_B(\tau_{\alpha_i},\tau_{\alpha_j})} 
\label{agmij} \\
& &  - \sum_{k,l=1}^m
[-\tau_1^{(i)}\delta_{ik} + \tau_2^{(j)}\delta_{jk}
  +X_k(\tau_{\alpha_i},\tau_{\alpha_j})] A^{-1}_{kl} 
[-\tau_1^{(i)}\delta_{il} + \tau_2^{(j)}\delta_{jl}
  +X_l(\tau_{\alpha_i},\tau_{\alpha_j})]\ , \nonumber
\eeqa
where we introduced the shorthand notation 
\beq
X_j(\tau_a,\tau_b) \equiv 
X(\tau_a,\tau_b;\tau_{\alpha_j},\tau_{\beta_j}).  
\label{short2}
\eeq
In the following section we will show how the string Green function 
reduces, in the relevant field theory limit, to the Green functions 
\eq{no7},\eq{agmii},\eq{agmi0} and \eq{agmij}.
\section{Reducing String Theory to Particle Theory}
\setcounter{equation}{0}
\indent

In this section we show how the string Green function, 
supplemented with our prescribed choice of local coordinates, 
correctly reproduces the particle theory Green functions 
\eq{no7},\eq{agmii},\eq{agmi0} and \eq{agmij} in the respective 
field theory limits.

The Green functions $G_{\mu \nu}^{(m)}$ correspond to different 
(classes of) $\Phi^3$-diagrams, depending on the values taken by 
$\mu,\nu\in \{0,1,\ldots,m\}$, and as explained in section 2 these 
different classes of diagrams correspond to different (classes of)
corners of string moduli space, hence to different classes of
pinching limits. But before we enter into the details of individual 
pinching limits, it is advantageous to rewrite the string Green 
function \eq{no8} on a form more similar to the world-line Green 
functions. As it stands, the string formula involves the period 
matrix, which is an $(m+1)\times (m+1)$-matrix, whereas the 
world-line formulae involve the $m \times m$-matrix $A$ and are 
expressed in terms of the one-loop Green function. 

To mimic the world-line formulation, where the particle diagram is 
constructed from a fundamental loop with total Schwinger proper time 
$T$, we single out the string loop labelled by $\mu=0$ and make the 
identification (as in eq.~\eq{deltai})
\beq
T=-{\alpha'\over2}\ln\vert k_0\vert =
{\alpha'\over2}(2\pi\mbox{Im}\tau)_{00}
\equiv {\alpha'\over2}\Delta \ . \label{no11}
\eeq
Our task is then to reexpress the $(m+1)$-loop Green function given 
by eq.~\eq{no8}, in terms of the one-loop Green function obtained by 
setting $m=0$ in eq.~\eq{no8}. In our standard configuration
\beq
z_{\alpha_0}=\infty, \hskip 20pt  z_{\beta_0}=0 \ ,      
\label{no14}
\eeq
we find
\beq
G^{(0)}_{\rm str} (z_1,z_2) = 
\ln\left\vert{z_1-z_2\over z^{1/2}_1 
z^{1/2}_2}\right\vert 
     -{1\over2\Delta}\ln^2\left\vert{z_2\over 
z_1}\right\vert
 + {\cal O} (k_{\mu},\bar{k}_{\mu}) \ .
\label{no13}
\eeq
Here we used our choice \eq{choice} of the local coordinate, which 
for the case of a torus in the standard configuration 
\eq{no14} is reduced to
\beq
        V_i'(0) = z_i \ . 
\eeq
This coincides with the choice made in ref.~\cite{paoloetal}.

We now turn our attention to the $(m+1)$-loop Green function given 
by eq.~\eq{no8}. As a first step we may rewrite it as 
follows~\footnote{ 
From now on we will stop writing explicitly the ${\cal O} (k_{\mu})$ 
terms, with the understanding that everything holds only 
in the limit of vanishing multipliers.}
\begin{eqnarray}
\lefteqn{G^{(m)}_{\rm str}(z_1,z_2) \ = \ \ln \left| 
\frac{z_1^{1/2} 
z_2^{1/2}}{(V_1'(0))^{1/2} (V_2'(0))^{1/2}} \right| +
G^{(0)}_{\rm str} (z_1,z_2)} \label{no16} \\
&&  + {1\over2} (\Omega_0)^2[ {1\over\Delta} - 
\imtau{00} ] 
- {1\over2}\sum_{i,j=1}^m\Omega_i\Omega_j\imtau{ij}
            - \Omega_0
            \sum_{i=1}^m\Omega_i\imtau{0i} \ ,           
   \nonumber 
\end{eqnarray}                                                    
where we used eq.~\eq{no13}.

In eq.~\eq{no16} the quantities $V_i'(0)$ refer to the choice of 
local coordinates on the genus $m+1$ Riemann surface and as such 
will be chosen differently in the various corners of moduli 
space, as explained in subsection 2.1.

Now, since the matrix \eq{no9} satisfies the obvious identity
\beq
\sum_{\mu=0}^m (2\pi \mbox{Im} \tau )_{0\mu} 
(2\pi \mbox{Im} \tau )^{-1}_{\mu \nu} =
\Delta \, 
(2\pi \mbox{Im} \tau)^{-1}_{0\nu} 
+ \sum_{i=1}^m x_i \, \imtau{i\nu} = \delta_{0\nu} 
\label{no17} \ ,
\eeq
where we defined
\beq
x_i \equiv \ln\left\vert{z_{\alpha_i}\over 
z_{\beta_i}}\right\vert = (2\pi \mbox{Im}\tau)_{0i} \ , 
\eeq
we can rewrite the 3rd and 5th terms on the r.h.s. of 
eq.~\eq{no16} as follows;
\beq
(2\pi \mbox{Im} \tau)^{-1}_{0i} 
= -\sum_{j=1}^m x_j \Delta^{-1} \imtau{ij} \ ,  
\label{no18}
\eeq
\beq
\frac{1}{\Delta} - (2\pi \mbox{Im} \tau)^{-1}_{00} 
= \sum_{i=1}^m x_i \Delta^{-1} (2\pi 
\mbox{Im}\tau )^{-1}_{i0} =
-\sum_{i,j=1}^m x_i x_j \Delta^{-2} \imtau{ij} \ ,  
\label{no19} 
\eeq
and thereby obtain
\begin{eqnarray}
G^{(m)}_{\rm str}(z_1,z_2) & = &
\ln \left| \frac{z_1^{1/2} 
z_2^{1/2}}{(V_1'(0))^{1/2} (V_2'(0))^{1/2}} \right| +
G^{(0)}_{\rm str} (z_1,z_2) \label{no20} \\
& & -{1\over2}\Delta^{-2}\sum_{i,j=1}^m
     (\Delta\Omega_i-x_i \Omega_0)(\Delta\Omega_j-x_j 
\Omega_0) \imtau{ij} \ .  
\nonumber
\end{eqnarray}                                                    
It is straightforward to verify that
\begin{eqnarray}
X_{\rm str} (z_1,z_2;\zai,\zbi) & \equiv & G^{(0)}_{\rm str} 
(z_1,\zai) - G^{(0)}_{\rm str} (z_1,\zbi) - G^{(0)}_{\rm str} 
(z_2,\zai) + G^{(0)}_{\rm str} (z_2,\zbi) \nonumber  \\
& = & \ln \left| \frac{z_2-\zbi}{z_1-\zbi} 
\frac{z_1-\zai}{z_2-\zai} \right| + \ln \left| \frac{\zai}{\zbi} 
\right| \ln \left| 
\frac{z_1}{z_2} \right| \Delta^{-1} \nonumber \\
& = & \Omega_i - x_i \Omega_0 \Delta^{-1} \ . 
\label{Xstring}
\end{eqnarray}
Furthermore, the $m \times m$ matrix defined by
\beq
A^{\rm str}_{ij} \equiv (2\pi \mbox{Im} \tau )_{ij} - 
x_i x_j \Delta^{-1} \ \ ; \ i,j = 1,\ldots, m \ , \eeq
is the inverse of the matrix 
$\{ (2\pi \mbox{Im} \tau)^{-1}_{ij} \}_{i,j=1,\ldots,m}$, 
i.e. satisfies
\beq
\sum_{k=1}^m A_{ik}^{\rm str} \, (2\pi {\rm Im} 
\tau)^{-1}_{kj} \ = \ 
\sum_{k=1}^m (2\pi {\rm Im} \tau)^{-1}_{ik} \, 
A_{kj}^{\rm str} \ = \ \delta_{ij} \ ,
\label{aidentity}
\eeq
as can be seen by inspection, using eq.~\eq{no18}.
By using eq.~\eq{Xstring}, replacing $z_1$ and $z_2$ 
with $\zaj$ and $\zbj$ respectively, we immediately find
\beq
A^{\rm str}_{ij} = -X_{\rm str} (\zai,\zbi;\zaj,\zbj) 
\ \ \mbox{for} \ \ i \neq j \ , \label{no31b}
\eeq
while for $i = j$ we have
\beq
A^{\rm str}_{ii} = -\ln|k_i| - \ln^2 \left| 
\frac{\zai}{\zbi} \right| 
\Delta^{-1} = \bar{\Delta}_i + 2G^{(0)}_{\rm str} (\zai,\zbi) \ , 
\label{no32b}
\eeq
where we defined
\beq
\bar{\Delta}_i \equiv -\ln \left\vert k_i 
\frac{(\zai-\zbi)^2}{\zai \zbi} 
\right\vert \ . \label{bardeltai}
\eeq
Collecting our results \eq{Xstring} and \eq{aidentity}, 
the world-sheet Green function \eq{no20} can now 
be written on the form
\begin{eqnarray}
G^{(m)}_{\rm str} (z_1,z_2) & = & 
\ln \left| \frac{z_1^{1/2} 
z_2^{1/2}}{(V_1'(0))^{1/2} (V_2'(0))^{1/2}} \right| +
G^{(0)}_{\rm str} (z_1,z_2) \label{green} \\
& & -\frac{1}{2} \sum_{i,j=1}^m X_{\rm str} 
(z_1,z_2;\zai,\zbi) 
X_{\rm str} (z_1,z_2;\zaj,\zbj) (A^{\rm str})^{-1}_{ij} 
\ . \nonumber 
\end{eqnarray}
We want to evaluate this expression in the pinching limits 
pertaining to the specific world-line Green functions 
$G_{00}^{(m)}$, $G_{ii}^{(m)}$, $G_{0i}^{(m)}$ and $G_{ij}^{(m)}$. 
As previously explained, each of these worldline Green functions 
corresponds to different ways of inserting two external legs in what 
is actually a whole class of $(m+1)$-loop $\Phi^3$ vacuum diagrams, 
namely any diagram that can be obtained by joining together $2m$ 
vertices located on a fundamental loop by means of $m$ internal 
propagators. Different $\Phi^3$ vacuum diagrams arise from choosing 
different orderings of the SPTs of the $2m$ vertices.

In the string formulation, the mapping of the parameters 
$\zai,\zbi$ into the local SPTs $\tau_{\alpha_i},\tau_{\beta_i}$ 
is straightforward, following the rules of subsection 2.3:
\begin{eqnarray}
\tau_{\alpha_i} & = & -{\alpha'\over2}\ln\vert \zai 
\vert \ \ \mbox{for} 
\ i=1,\ldots, m \ ,  \label{no12} \\
\tau_{\beta_i} & = & -{\alpha'\over2}\ln\vert \zbi \vert 
\ \ \mbox{for} 
\ i=1,\ldots, m \ . \nonumber 
\end{eqnarray}
Different orderings of the local proper times imply different 
pinching limits for the parameters $\zai$ and $\zbi$. 
For example the ordering
$$
\tau_{\alpha_1} < \tau_{\alpha_2} < \dots < 
\tau_{\alpha_m} <  
\tau_{\beta_m} < \dots < \tau_{\beta_1} 
$$
is equivalent to the pinching limit
$$
\vert z_{\beta_1} \vert \ll \dots \ll \vert z_{\beta_m} 
\vert \ll
\vert z_{\alpha_m} \vert \ll 
\dots \ll \vert z_{\alpha_1} \vert 
$$
in the limit $\alpha' \ra 0$.

We can have $(2m)!$ different orderings of these local proper 
times, and therefore as many different pinching limits. To analyze 
the field theory limit case by case is possible (we did it), 
but very tedious. Fortunately, there is a much quicker way:
The Green function, as written in eq.~\eq{green}, is expressed 
solely in terms of the one-loop Green function \eq{no13}, and it is 
therefore sufficient to consider the field theory limit of this 
quantity. It is easy to verify that for any two points $z_a$ and 
$z_b$ related to SPTs $\tau_a$ and $\tau_b$ by eq.~\eq{no12} we have
\beq
G^{(0)}_{\rm str} (z_a,z_b) \ \alimit \ 
\frac{1}{\alpha'} G_B 
(\tau_a,\tau_b) \ , \label{onelooppinch}
\eeq
regardless of which one of the two 
possible orderings of $\tau_a$ and $\tau_b$ we happen to consider.
The proof is closely analogous to the one given for the 
cylinder in subsection 2.1.

In view of the scaling
behaviour \eq{onelooppinch} we see that for $i \neq j$
\begin{eqnarray}
\lefteqn{A_{ij}^{\rm str} = -X_{\rm str} 
(\zai,\zbi;\zaj,\zbj) \ 
\alimit} \label{aijlim} \\
& & -\frac{1}{\alpha'} \left( 
G_B(\tau_{\alpha_i},\tau_{\alpha_j})
- G_B(\tau_{\alpha_i},\tau_{\beta_j}) - 
G_B(\tau_{\beta_i},\tau_{\alpha_j})
+ G_B(\tau_{\beta_i},\tau_{\beta_j}) \right) \ = 
\nonumber \\
& & - \frac{2}{\alpha'} 
X(\tau_{\alpha_i},\tau_{\beta_i};
\tau_{\alpha_j},\tau_{\beta_j}) = \frac{2}{\alpha'} 
A_{ij} \ ,
\end{eqnarray}
where we used eqs.~\eq{no31b}, \eq{Xstring}, \eq{no5} and 
\eq{no4}.

In the diagonal case $i=j$ we encounter the quantity 
$\bar{\Delta}_i$, defined by eq.~\eq{bardeltai}, which has the 
following behaviour
\beq
\bar{\Delta}_i \ \alimit \ \frac{2}{\alpha'} \left( T_i - \vert 
\tau_{\alpha_i} - \tau_{\beta_i} \vert \right) = 
\frac{2}{\alpha'} \bar{T}_i \ . \label{deltailim}
\eeq
Here we see the fact that $T_i$, defined by eq.~\eq{deltai}, is 
the SPT of the entire loop formed by the insertion of the 
internal propagator, whereas $\bar{T}_i$ is only the SPT of the 
internal propagator itself.

Using the behaviour \eq{deltailim} and \eq{onelooppinch} in 
eq.~\eq{no32b}, we find
\beq
A_{ii}^{\rm str} \ \alimit \ \frac{2}{\alpha'} \left( 
\bar{T}_i + G_B (\tau_{\alpha_i},\tau_{\beta_i}) \right) = 
\frac{2}{\alpha'} A_{ii} \ . 
\label{aiilim}
\eeq
In summary we have the scaling relation
\beq
A_{ij}^{\rm str} \ \alimit \ \frac{2}{\alpha'} A_{ij} 
\label{aijlimtwo}
\eeq
for off-diagonal as well as diagonal elements.

To investigate the scaling behaviour of the other quantities 
appearing in the string Green function \eq{green} we need to 
specify the pinching limit for $z_1$ and $z_2$; i.e., we have 
to consider case by case the situations corresponding to the 
various particle Green functions.
\subsection{The Green Function $G_{00}^{(m)}$}
\indent

The simplest case is when both external legs are inserted on the 
fundamental loop (Fig.3). In this case the tree diagram obtained 
by cutting open all the loops consists only of one (main) branch 
and the mapping into SPTs is given by eq.~\eq{mainspt}, i.e.
\beq
\tau_i = - \frac{\alpha'}{2} \ln \vert z_i \vert \ \ 
\mbox{for} \ \ i =1,2 \ . 
\eeq
Thus the scaling behaviour \eq{onelooppinch} also holds for Green 
functions having $z_1$ and/or $z_2$ as argument. Accordingly
\begin{eqnarray}
G_{\rm str}^{(0)} (z_1,z_2) & \alimit & 
\frac{1}{\alpha'} G_B 
(\tau_1,\tau_2) \label{lim00} \\
X_{\rm str} (z_1,z_2; \zai,\zbi) & \alimit & 
\frac{2}{\alpha'} 
X(\tau_1,\tau_2;\tau_{\alpha_i},\tau_{\beta_i}) \ .  
\nonumber
\end{eqnarray}
\begin{figure}
\begin{center}
\input{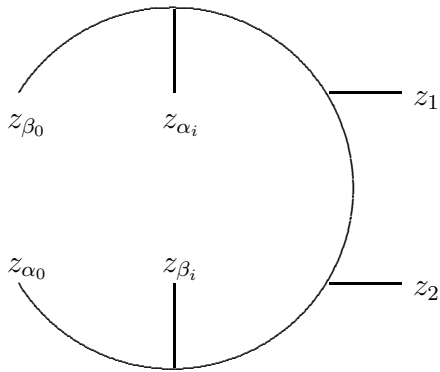}
\end{center}
\caption[gyhhh]{This is a typical representative of the 
class of $\Phi^3$ diagrams relevant for the Green function 
$G_{00}^{(m)}$. It corresponds to a particular ordering of the 
legs $z_1$, $z_2$, $z_{\alpha_i}$ and $z_{\beta_i}$. 
The other legs (labelled by $z_{\alpha_j}$ and 
$z_{\beta_j}$, $j \neq i$) have been suppressed.}
\end{figure}
Finally, for $V_i'(0)$ we should use eq.~\eq{omegachoice} where 
for $\omega$ we may use any one of the holomorphic one-forms 
that satisfies the normalization condition \eq{normaliz}. The 
homology cycle $a$ is represented in the complex plane by a contour 
that encircles $z_{\beta_0}=0$ as well as all those points $\zaj$ 
and $\zbj$that label vertices with SPT values larger than $\tau_i$. 
Thus, if we consider the standard basis of 
holomorphic one-forms, where
\beq
\omega_{\mu} (z) = \frac{z_{\beta_{\mu}} - 
z_{\alpha_{\mu}}}{(z-z_{\beta_{\mu}})(z-z_{\alpha_{\mu}})} \ ; \ \ 
\mu=0,1,\ldots,m \ , \label{oneformbasis} 
\eeq
we see that, apart from $\omega_0$, only those $\omega_j$, 
$j=1,\ldots,m$,  for which either
\beq
\vert z_{\beta_{j}} \vert \ll \vert z_i \vert \ll \vert 
z_{\alpha_{j}} \vert \ \ \ \ \mbox{or} \ \ \ \
\vert z_{\alpha_{j}} \vert \ll \vert z_i \vert \ll \vert 
z_{\beta_{j}} \vert \label{doublelim}
\eeq
will satisfy the normalization condition \eq{normaliz}. The other 
one-forms will integrate to zero, since the contour $a$ 
encircles either both or none of the singularities of $\omega_{j}$. 
In either of the cases \eq{doublelim} we find
\beq
\vert \omega_{\mu} (z_i) \vert \simeq \vert z_i \vert^{-1} \ , 
\eeq
which explicitly shows how all choices of holomorphic one-forms 
subject to the normalization condition \eq{normaliz} are 
equivalent in the limit $\alpha' \rightarrow 0$. In short, our 
prescription for the local coordinate reduces in this case to
\beq
\vert V_i'(0) \vert = \vert z_i \vert \ , \ i=1,2 \ . 
\label{choice00}
\eeq
Using the scaling relations \eq{aijlimtwo} and \eq{lim00} in 
eq.~\eq{green}, together with the choice \eq{choice00}, we 
readily recover the particle Green function
\beq
G^{(m)}_{\rm str} (z_1,z_2) \alimit \ \frac{1}{\alpha'} 
G_{00}^{(m)} 
(\tau_1,\tau_2) = \frac{1}{\alpha'} 
\tilde{G}_B^{(m)}(\tau_1,\tau_2) \ .
\eeq
\subsection{The Green Function $G_{ii}^{(m)}$}
\indent

In this case the tree diagram obtained by cutting open all loops 
consists of the main branch plus the side-branch where the two 
external legs are inserted (Fig.4). If we choose to let the SPT 
flow $\tau^{(i)}$ end at the leg labelled by $z_{\alpha_i}$, 
we have, according to eq.~\eq{SPTabs}
\beq
\tau_j^{(i)} = - \frac{\alpha'}{2} \ln \left\vert \frac{z_j - 
z_{\alpha_i}}{z_{\alpha_i}} \right\vert \ \ \mbox{for} \ 
j=1,2 \ .
\eeq
Here $\tau^{(i)} = 0$ corresponds to the point 
$\tau=\tau_{\alpha_i}$ on the fundamental loop, whereas 
$\tau^{(i)} = \bar{T}_i$ 
corresponds to the point $\tau = \tau_{\beta_i}$, in 
accordance with the conventions~\footnote{It is, of course, possible 
to adopt the reverse notation, which corresponds to the 
interchange of $\alpha_i\leftrightarrow\beta_i$. We verified 
that the relation between string and particle Green 
functions remains unchanged, as long as a similar change of 
convention is adopted
in the particle formulae eqs.~\eq{agmi0}, \eq{agmij}. } 
underlying the particle Green functions 
of section 3. We then have the pinching limit
\begin{eqnarray}
\vert z_1-z_{\alpha_i} \vert \ll \vert z_2 - 
z_{\alpha_i} \vert \ll 
\vert z_{\alpha_i} \vert & \mbox{if} & \tau^{(i)}_1 > 
\tau^{(i)}_2 
\label{doublelimii} \\
\vert z_2-z_{\alpha_i} \vert \ll \vert z_1 - 
z_{\alpha_i} \vert \ll 
\vert z_{\alpha_i} \vert & \mbox{if} & \tau^{(i)}_2 > 
\tau^{(i)}_1 \ .
\nonumber
\end{eqnarray}
\begin{figure}
\begin{center}
\input{multifig4}
\end{center}
\caption[hyu]{The typical $\Phi^3$ diagram relevant for the Green 
function $G_{ii}^{(m)}$. The legs labelled by $z_{\alpha_j}$ and 
$z_{\beta_j}$, $j \neq i$, have been suppressed.}
\end{figure}
In this case the homology cycle $a$ is simply equal to 
the basis cycle $a_i$ in the homology basis dual to the basis 
\eq{oneformbasis} of holomorphic one-forms, and accordingly 
we may take
\beq
\left| V_j'(0) \right| = \left| \omega_i (z_j) 
\right|^{-1} = \left|
\frac{(z_j-z_{\alpha_i})(z_j-z_{\beta_i})}{z_{\alpha_i}-
z_{\beta_i}} 
\right| \simeq \left| z_j - z_{\alpha_i} \right| \ 
\mbox{for} \ j=1,2 \ 
.
\eeq
It is now straightforward to verify that in the pinching 
limit \eq{doublelimii}
\begin{eqnarray}
\lefteqn{\ln \left| \frac{z_1^{1/2} 
z_2^{1/2}}{(V_1'(0))^{1/2} (V_2'(0))^{1/2}} 
\right| + G^{(0)}_{\rm str} (z_1,z_2) \simeq } 
\label{limiione}  \\  
& & \ln 
\left|\frac{(z_1-z_{\alpha_i})-(z_2-z_{\alpha_i})}{(z_1-
z_{\alpha_i})^{1/2}
(z_2-z_{\alpha_i})^{1/2}}\right| \ \alimit \ 
\frac{1}{\alpha'} \vert 
\tau^{(i)}_1 - \tau_2^{(i)} \vert \ , \nonumber
\end{eqnarray}
while for $j=1,2$ and $k=1,\ldots,m$
\begin{eqnarray}
G_{\rm str}^{(0)}(z_j,z_{\beta_k}) & \alimit & 
\frac{1}{\alpha'} G_B 
(\tau_{\alpha_i},\tau_{\beta_k}) \ , \nonumber \\
G_{\rm str}^{(0)}(z_j,z_{\alpha_k}) & \alimit & 
\frac{1}{\alpha'}
\left\{ G_B (\tau_{\alpha_i},\tau_{\alpha_k}) 
(1-\delta_{ik}) - 2 \tau_j^{(i)} 
\delta_{ik} \right\} \ , \nonumber
\end{eqnarray}
so that
\beq
X_{\rm str} (z_1,z_2;z_{\alpha_k},z_{\beta_k}) \ \alimit \ - 
\frac{2}{\alpha'} (\tau_1^{(i)}-\tau_2^{(i)}) 
\delta_{ik} \ . 
\label{limiitwo}
\eeq
Combining eqs.\eq{limiione}, \eq{limiitwo} and 
\eq{aijlimtwo} we find
\beq
G_{\rm str}^{(m)} (z_1,z_2) \ \alimit \ 
\frac{1}{\alpha'} G^{(m)}_{ii} 
(\tau_1^{(i)},\tau_2^{(i)}) \ , 
\eeq
as desired.
\subsection{The Green Function $G_{i0}^{(m)}$}
\indent

This case is a mixture of the previous two (Fig.5). We 
have
\begin{equation}
\begin{array}{lll}
\tau_1^{(i)} & = & \displaystyle{- \frac{\alpha'}{2} \ln 
\left| \frac{z_1-z_{\alpha_i}}{z_{\alpha_i}} \right|} \\
\tau_2 & = & \displaystyle{- \frac{\alpha'}{2} \ln \vert 
z_2 \vert }
\end{array} \ \ \ \ \ \ \ \ 
\begin{array}{lll}
\vert V_1'(0) \vert & = & \vert z_1 - z_{\alpha_i} \vert 
\\
\vert V_2'(0) \vert & = & \vert z_2 \vert \ ,
\end{array}  
\end{equation}
and in the relevant pinching limit 
($\vert z_2 \vert \ll 1 $ and $ |z_1 
- z_{\alpha_i} | \ll |z_{\alpha_i} | \ll 1$) we find
\begin{figure}
\begin{center}
\input{multifig5}
\end{center}
\caption[hytr]{The typical $\Phi^3$ diagram relevant for the Green 
function $G_{i0}^{(m)}$. The legs labelled by $z_{\alpha_j}$ 
and $z_{\beta_j}$, $j \neq i$, have been suppressed.}
\end{figure}
\begin{equation}
\ln \left| \frac{z_1^{1/2} z_2^{1/2}}{(V_1'(0))^{1/2} 
(V_2'(0))^{1/2}} 
\right| + G^{(0)}_{\rm str} (z_1,z_2) \ \alimit \  
\frac{1}{\alpha'} \left[ G_B(\tau_{\alpha_i},\tau_2) + 
\tau_1^{(i)} \right] 
\end{equation}
\beq
X_{\rm str} (z_1,z_2;z_{\alpha_j},z_{\beta_j}) \ \alimit \
\frac{2}{\alpha'} \left[ 
X(\tau_{\alpha_i},\tau_2;\tau_{\alpha_j},
\tau_{\beta_j}) - \tau_1^{(i)} 
\delta_{ij} \right] \ , 
\eeq
which, combined with eq.~\eq{aijlimtwo}, leads to the 
correct result, i.e. that
\beq
G_{\rm str}^{(m)} (z_1,z_2) \ \alimit \ 
\frac{1}{\alpha'} G_{i0}^{(m)} 
(\tau_1^{(i)},\tau_2) \ . 
\eeq
\subsection{The Green Function $G_{ij}^{(m)}$ ($i \neq j$)}
\indent

In this case, when we cut open the loops of the $\Phi^3$-diagram, 
we have two side-branches inserted on the main branch: one 
ending at $z_{\alpha_i}$, with a stem labelled by $z_1$; and 
another ending at $z_{\alpha_j}$, with a stem labelled by $z_2$ 
(Fig.6). The mapping into SPTs and the choice of local coordinates 
are as follows
\begin{equation}
\begin{array}{lll}
\tau_1^{(i)} & = & \displaystyle{- \frac{\alpha'}{2} \ln 
\left| \frac{z_1-z_{\alpha_i}}{z_{\alpha_i}} \right|}  \\
\tau_2^{(j)} & = & \displaystyle{- \frac{\alpha'}{2} \ln 
\left| \frac{z_2-z_{\alpha_j}}{z_{\alpha_j}} \right| }
\end{array}
 \ \ \ \ \ \
\begin{array}{lll}
\vert V_1'(0) \vert & = & \vert z_1 -z_{\alpha_i} \vert  \\
\vert V_2'(0) \vert & = & \vert z_2 -z_{\alpha_j} \vert \ , 
\end{array} 
\end{equation}
\begin{figure}
\begin{center}
\input{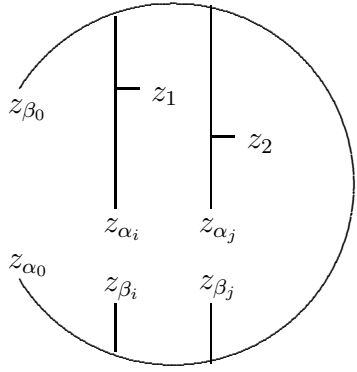}
\end{center}
\caption[loi]{The typical $\Phi^3$ diagram relevant for 
the Green function $G_{ij}^{(m)}$.
The legs labelled by $z_{\alpha_k}$ and 
$z_{\beta_k}$, $k \neq i,j$, have been suppressed.}
\end{figure}
and in the pinching limit 
$\vert z_1 - z_{\alpha_i} \vert \ll \vert 
z_{\alpha_i} \vert $ and $\vert z_2 - z_{\alpha_j} \vert 
\ll \vert z_{\alpha_j} \vert $ we find
\begin{equation}
\ln \left| \frac{z_1^{1/2} z_2^{1/2}}{(V_1'(0))^{1/2} 
(V_2'(0))^{1/2}} 
\right| + G^{(0)}_{\rm str} (z_1,z_2) \ \alimit \ 
\frac{1}{\alpha'} \left[ 
G_B(\tau_{\alpha_i},\tau_{\alpha_j}) + 
\tau_1^{(i)} + \tau_2^{(j)} 
\right] 
\end{equation}
\beq
X_{\rm str} (z_1,z_2;z_{\alpha_k},z_{\beta_k}) \ \alimit 
\
\frac{2}{\alpha'} \left[ 
X(\tau_{\alpha_i},\tau_{\alpha_j};
\tau_{\alpha_k},\tau_{\beta_k}) 
- \tau_1^{(i)} \delta_{ik} + \tau_2^{(j)} \delta_{jk} 
\right] \ , 
\eeq
and thus, as expected
\beq
G_{\rm str}^{(m)} (z_1,z_2) \ \alimit \ 
\frac{1}{\alpha'} G_{ij}^{(m)} 
(\tau_1^{(i)},\tau_2^{(j)}) \ . 
\eeq
\section{Conclusions and Open Problems}
\setcounter{equation}{0}
\indent

We have described how the particle Green function 
(corresponding to any given $\Phi^3$ vacuum diagram) can 
be obtained from the string Green function by approaching the 
corner in string moduli space where the string world sheet 
degenerates into the desired particle diagram, more precisely, 
by taking the limit $\alpha' \ra 0$, keeping all Schwinger Proper 
Times of the particle diagram fixed at finite values. 

Since the string Green function is not conformally invariant, one
will only recover the particle Green function in this limit if  
an appropriate choice is made for the set of local coordinates, 
more precisely, for the quantities $\vert V_i'(0) \vert$, in the 
limit of degenerate string world sheets. We have 
prescribed how to make this choice.

We have also given a set of simple rules, formulated in the 
explicit setting of the Schottky parametrization, that allows 
one to identify the corner of string moduli space corresponding to 
any given $N$-point multiloop $\Phi^3$ particle diagram, 
as well as the precise mapping of string moduli into the 
Schwinger Proper Times pertaining to such a diagram. 

Our general procedure has been verified by explicit comparison with 
all types of particle world-line Green functions corresponding 
to the Schmidt-Schubert class of $\Phi^3$ vacuum diagrams.

Although we have chosen to express the world-line Green functions 
in terms of the SPT parameters $T$ of the fundamental loop 
and ${\bar T}_i$ of the internal propagators, generally speaking 
there is no unique way of introducing the SPT parametrization 
for a given particle diagram. However, the prescription 
we have given for taking the field theory limit is entirely 
geometrical and therefore should not depend on the choice of SPT 
parametrization. Accordingly, although a different choice of SPT 
parametrization will lead to a new world-line Green function, this 
new Green function should be related to the old one merely by the 
appropriate change of SPT variables, as discussed in the two-loop 
case in ref.~\cite{sato}.

Understanding how the bosonic world-line Green function of particle 
theory is recovered from the world-sheet Green function of string 
theory is obviously a relevant step towards the formulation of 
string-based rules for multiloop amplitudes in {\em any} field 
theory.

In the case of $\Phi^3$ theory, analyzed from the particle point 
of view in ref.~\cite{SSphi} and from the string point of view 
in ref.~\cite{Paolonew}, it constitutes the major part of 
such a formulation.

However, in the physically more interesting case of 
Yang-Mills theory, various obstacles remain. First of all, the 
structure of the $3$-point interactions is much more complicated 
than in $\Phi^3$ theory. In string theory this new structure of the 
scattering amplitudes arises in various ways: In the bosonic 
string~\cite{Zviletter},\cite{paoloetal} it appears as a result of 
having to expand the modular integrand around the tachyon poles in 
order to obtain the massless poles. 
In the superstring~\cite{BK}, it appears as a result of having also 
fermionic fields on the world-sheet.
In either case, some severe problems arise at multiloop level.
In the bosonic string, the subtraction of the tachyon 
poles introduces potential ambiguities in the remaining massless 
amplitude~\cite{Zviandme}. In the superstring one has to 
deal with either an integration over supermoduli~\cite{me} or the 
insertion of picture changing operators~\cite{me2}. Either way, one 
will face the problem of total derivatives, making the explicit 
analysis potentially ambiguous and at any rate very delicate.

Another complication of Yang-Mills theory as compared to $\Phi^3$ 
theory is the existence of contact terms. At one-loop level all 
contact terms may be removed by a partial integration with 
respect to the Koba-Nielsen variables~\cite{BK}, 
but, as argued in detail in ref.~\cite{me2}, this does not seem to 
be possible at multiloop level. Contact term contributions 
to the amplitude can arise from string world sheets that are 
{\em not} fully pinched, i.e. where 
two vertices need not be separated by an (almost) infintely long 
cylinder. Therefore, to extract all contact terms from 
the string theory amplitude one needs to integrate not only over 
the various ``$\Phi^3$-like'' corners of moduli space that we have 
investigated in this paper, but also over appropriate parts of 
moduli space interpolating between these corners. 
A tractable way of doing this has yet to be found.
\appendix
\section{Proof of eqs.~(3.19) and (3.20)}
\setcounter{equation}{0}
\indent

In this Appendix we present a proof of the two 
identities \eq{app1} and \eq{app2} quoted in section 3.
We start from the defining relation of $G_B^{(m)}$ 
(given by eq.(16) in ref.~[6])
\beq
G_B^{(m)}(\tau_1,\tau_2)=G_B(\tau_1,\tau_2)+{1\over2}
\sum_{k=1}^m 
{\bar T}_k^{-1}g^{(m)}_k(\tau_1)g_k(\tau_2) \ , 
\label{defofgb}
\eeq
where
\beq
g^{(m)}_k(\tau) \equiv G_B^{(m)}(\tau,\tau_{\alpha_k})-
G_B^{(m)}(\tau,\tau_{\beta_k}) \ ,
\eeq
with the understanding that $g_k = g_k^{(0)}$ and 
$G_B^{(0)} = G_B$. The relation \eq{defofgb} is very useful for 
checking various algebraic identities concerning $G_B^{(m)}$, 
since it does not involve explicitly the inverse of the matrix $A$.

In addition, we list the following algebraic relations which can be 
shown directly from the definitions 
\eq{no46b}, \eq{short1}, \eq{no2} and \eq{no5}. 
\begin{equation}
X^{(m)}_{ij}=X^{(m)}_{ji}=X_{ij}+\sum_{k,l=1}^m
X_{ik}A^{-1}_{kl}X_{lj} \ ,
\end{equation}
\beq
X^{(m)}_{ii}=-{\tilde G}_B^{(m)}
(\tau_{\alpha_i},\tau_{\beta_i}) \ ,
\eeq
\beqa
X^{(m)}(a,b;c,d) & = & -X^{(m)}(b,a;c,d) = 
-X^{(m)}(a,b;d,c)\nn\\
                 & = & X^{(m)}(b,a;d,c) = 
X^{(m)}(c,d;a,b) \ .
\eeqa

Now, applying eq.~\eq{defofgb} to the definition 
\eq{no46b} we obtain
\beq
X^{(m)}(\tau_a,\tau_b;\tau_c,\tau_d)=X(\tau_a,\tau_b;
\tau_c,\tau_d) + \sum_{k=1}^m{\bar 
T}_k^{-1}X^{(m)}_k(\tau_a,\tau_b)X_k(\tau_c,\tau_d) \ .
\label{defofxm}
\eeq
Since both $X^{(m)}$ and $X$ are symmetric under the interchange
$(\tau_a,\tau_b) \leftrightarrow (\tau_c,\tau_d)$, so is 
the second term on the right-hand side of eq.~\eq{defofxm}. 
This allows us to write eq.~\eq{defofxm} on the alternative form
\beq
X^{(m)}(\tau_a,\tau_b;\tau_c,\tau_d)=X(\tau_a,\tau_b;
\tau_c,\tau_d)+
\sum_{k=1}^m{\bar T}_k^{-1} X_k(\tau_a,\tau_b) 
X^{(m)}_k(\tau_c,\tau_d) \ .
\label{defofxmb}
\eeq
We are now in a position to verify the formula for the 
inverse matrix, $A^{-1}$, given by eq.~\eq{app1}. Multiplying the 
right-hand side of this equation by $A$, as defined in 
eq.~\eq{no4}, and using eq.~\eq{defofxm}, we immediately find
\[
\sum_{k=1}^m {1 \over \bar{T}_i \bar{T}_k} \left( 
\bar{T}_i \delta_{ik} 
+ X^{(m)}_{ik} \right) A_{kj} \ = \ 
\delta_{ij}+{1\over{\bar T}_i}\left(
X^{(m)}_{ij}-X_{ij}-\sum_{k=1}^m{1\over{\bar T}_k}
X^{(m)}_{ik}X_{kj} \right)=\delta_{ij} \ .
\]
The notations $X_{ij}$ and $X_k$ are explained in 
eqs.~\eq{short1} and \eq{short2}.

Next, to prove eq.~\eq{app2} we use eq.~\eq{defofxmb}, 
according to which
\beqa
X^{(m)}(\tau_a,\tau_b;\tau_{\alpha_i},\tau_{\beta_i}) & = & 
X(\tau_a,\tau_b;\tau_{\alpha_i},\tau_{\beta_i}) +
\sum_{k=1}^m{1\over{\bar T}_k} 
X^{(m)}_{k}(\tau_{\alpha_i},\tau_{\beta_i}) 
X_k(\tau_a,\tau_b) \nn \\
& = & \sum_{k=1}^m({1\over{\bar 
T}_k}X^{(m)}_{ik}+\delta_{ik}) 
      X_k(\tau_a,\tau_b) \nn \\
& = & {\bar T}_i 
\sum_{k=1}^mA^{-1}_{ik}X_k(\tau_a,\tau_b) \ ,
\eeqa
where eq.~\eq{app1} is used at the bottom line. This 
proves eq.~\eq{app2}.

\newpage

%
\end{document}